\documentclass[12pt,epsf,axodraw]{article}

\newcommand{\bea}{\begin{eqnarray}}
\newcommand{\eea}{\end{eqnarray}}
\newcommand{\be}{\begin{equation}}
\newcommand{\ee}{\begin{equation}}

\begin{document}

\thispagestyle{empty}

\begin{flushright}
{\tt  VEC/PHYSICS/P/1/2004-2005}
\end{flushright}

\begin{center}
{\LARGE \bf A hyperbolic parametrization of lepton mixing for small
$U_{e3}$\\}
\vspace{2cm}
{\bf Biswajoy Brahmachari}
\\
\vskip 1cm
{\sl Department of Physics, Vidyasagar Evening College,\\
39, Sankar Ghosh Lane, Kolkata 700006, India.\\}
\end{center}
\vskip 1cm
\begin{abstract}

We use hyperbolic functions to parametrize lepton mixing
matrix using only one input parameter $\phi$. This matrix has
three mixing angles as outputs therefore giving two predictions. In
particular it predicts $U_{e3}=0$ besides predicting correct solar
and atmospheric mixing. We confine us to real $\phi$.
For complex $\phi$ mixing matrix is no longer unitary.
Next we accentuate this unitary mixing matrix with an additional small
parameter $\delta$ while keeping unitarity of the matrix exact. When
this second parameter is included, with this framework, one can handle
small but non-zero $U_{e3}$. In the second case from two input parameters
one obtains three mixing angles thus one prediction.

\end{abstract}

\vskip 1in

\begin{flushleft}
{\tt email: biswajoy.brahmachari@cern.ch}
\end{flushleft}

\newpage

Heroic advancements in the realm of neutrino physics experiments
\cite{ref1-exp1,ref1-exp2,ref1-exp3,ref1-exp4,ref1-exp5,ref1-exp6,ref1-exp7}
have given us substantial information on lepton mixing angles. Many
different groups have extracted lepton mixing angles from raw data
\cite{ref2-data1,ref2-data2,ref2-data3,ref2-data4,ref2-data5,ref2-data6,ref2-data7}.
Using these informations magnitudes of nine elements of this $3
\times 3$ unitary mixing matrix $U_{lep}$ approximately looks like
\cite{ref3-unum1,ref3-unum2,ref3-unum3,ref3-unum4,ref3-unum5},
\begin{equation}
|U_{lep}| =\pmatrix{0.72-0.88 & 0.46-0.68 & < 0.22 \cr
0.25-0.65 & 0.27-0.73 & 0.55-0.84 \cr
0.10-0.57 & 0.41-0.80 & 0.52-0.83}. \label{para}
\end{equation}

Very little is known theoretically on how leptons perform their
mixing of generations. To construct a theory of flavor in a
ground-up approach one needs to first do a theoretical
parametrization of the extracted numbers and then search a theory
which can generate this pattern of mixing. Therefore we have to make
a few ansatze first in order to proceed toward a fuller theory. A
number of them already exists in the literature\cite{ref4-ansatz1,
ref4-ansatz2,ref4-ansatz3,ref4-ansatz4,ref4-ansatz5,ref4-ansatz6,ref4-ansatz7,
ref4-ansatz8,ref4-ansatz9,ref4-ansatz-9-1,ref4-ansatz10,ref4-ansatz11,ref4-ansatz12,ref4-ansatz13,ref4-ansatz14}.
We have cited a few examples in Table \ref{table1} to motivate the
present work.

\begin{table}[htb]
\begin{center}
\[
\begin{array}{|c c|}
\hline
&\cr
\underbar{The $\omega$ mixing}
&
U={ 1 \over \sqrt{3}} \pmatrix{1 & \omega & \omega^2 \cr
1 & \omega^2 & \omega \cr
1 & 1 & 1}
\cr
&\cr
\underbar{Bi-maximal}
&
U=\pmatrix{{ 1 \over \sqrt{2}} & -{ 1 \over \sqrt{2}} & 0 \cr
{1 \over 2} & { 1 \over 2} & -{ 1 \over \sqrt{2}} \cr
{ 1 \over 2} & { 1 \over 2} & { 1 \over \sqrt{2}}}
\cr
&\cr
\underbar{ Zee}
&
U=\pmatrix{
-{2 \over \sqrt{6}} & {1 \over \sqrt{3}} & 0 \cr
{ 1 \over \sqrt{6}} & { 1 \over \sqrt{3}} & { 1 \over \sqrt{2}}\cr
 { 1 \over \sqrt{6}} & { 1 \over \sqrt{3}} & -{ 1 \over \sqrt{2}}}
\cr
&\cr
\underbar{Giunti}
&
U=\pmatrix{-{\sqrt{3} \over 2} & { 1 \over 2} & 0 \cr
{ 1 \over 2 \sqrt{2}} & {\sqrt{3} \over 2 \sqrt{2}} & { 1 \over
\sqrt{2}} \cr
{ 1 \over 2 \sqrt{2}} & {\sqrt{3} \over 2 \sqrt{2}} & -{ 1 \over
\sqrt{2}} }
\cr
&\cr
\underbar{sine-cosine}
&
U=\pmatrix{\cos \theta & \sin \theta & 0 \cr
-{\sin \theta \over \sqrt{2}} & {\cos \theta \over \sqrt{2}} & { 1
\over \sqrt{2}} \cr
{\sin \theta \over \sqrt{2}} & -{\cos \theta \over \sqrt{2}} & { 1
\over \sqrt{2}}}
\cr
&\cr
\underbar{Tri-maximal} &
U={ 1 \over \sqrt{3}} \pmatrix{1 & 1 & 1 \cr
1 & 1 & 1 \cr
1 & 1 & 1}
\cr
&\cr
\underbar{Minakata-Yasuda/Fritzsch and Zing}
&
U=\pmatrix{{1 \over \sqrt{2}} & {i \over \sqrt{2}} & 0 \cr
{1 \over \sqrt{6}} & -{i \over \sqrt{6}} & -{ 2
\over \sqrt{6}} \cr
{1 \over \sqrt{3}} & -{i \over \sqrt{3}} & { 1
\over \sqrt{3}}}
\cr
&\cr
\hline
\end{array}
\]
\end{center}
\caption{A selection of a few ansatze of lepton mixing that
exist in literature. This selection is only a sample, not a complete list.}
\label{table1}
\end{table}

At this point let us briefly discuss mixing in the lepton sector and
fix the definition of the $U$ matrix in the context of a
$G \equiv SU(2)_L \times U(1)_Y$ model. Rotation in flavor space
becomes visible in interactions between charge changing weak currents
and a gauge boson (lets call mass eigenstates of gauge bosons $A_\mu$
and gauge eigenstates $A^0_\mu$) after $G$ symmetry is broken. In
the $G$ invariant form interactions of left handed gauge eigentates
is diagonal in flavor space, and looks as following;
\begin{equation}
(\overline{l}^\alpha_L \gamma^\mu \nu^\alpha_L)~A^0_\mu, \label{int1}
\end{equation}
where $A^0_\mu,l^\alpha_L,\nu^\alpha_L$ are gauge eigenstates. When
$SU(2)_L \times U(1)_Y$ symmetry is broken by the Higgs mechanism
or extra-dimensional mechanism or some other yet-unknown mechanism,
naturally gauge eigenstates no-longer exist, and the gauge bosons now
become massive. Interaction given in Eqn \ref{int1}, can now be easily
rearranged in terms of mass eigenstates as,
\begin{equation}
(\overline{l}^i \gamma^\mu U_{ij} \nu^j)~A_\mu. \label{int2}
\end{equation}
We have defined  unitary mixing matrix $U$ of lepton sector in this
conventional way given in Eqn. \ref{int2}. We do not commit to any
specific method of $SU(2)_L \times U(1)$ symmetry breaking and simply
concentrate on a parametrization of the mixing matrix of lepton sector
which will produce Eqn. \ref{para} in a natural way. If we
invoke a specific mechanism of symmetry breaking or a specific
flavor symmetry, present paper would be less generally valid and hence
weaker.

Now we are ready to put forward the first ansatz, in which the input
parameter $\phi$ may either be real or complex. When the input
parameter is real from one input we will get three mixing angles, and
when $\phi$ is complex, from two real inputs (a complex number can be
thought of two real numbers), we will predict three mixing angles and
the CP violating parameter $J$. In both cases, there are two
predictions. However we will presently see that complex values of
$\phi$ will lead to violation of unitarity of the matrix $U_0$ so the
predictions are not dependable for complex $\phi$. Therefore we will
confine ourselves to real $\phi$.

\noindent \underbar{HYPERBOLIC ANSATZ 1}

\begin{equation}
U^0 =\pmatrix{
    -{\rm sech}\ \phi & \tanh \phi & 0 \cr
   { \tanh \phi \over \sqrt{2}} &
   { {\rm sech}\ \phi \over \sqrt{2}} & { 1 \over \sqrt{2}} \cr
    { \tanh \phi \over \sqrt{2}} &
    { {\rm sech}\ \phi \over \sqrt{2}} &
    -{ 1 \over \sqrt{2}}} ~~~~|\phi| \approx {\pi / 6}.
\end{equation}

\begin{table}[ht]
\begin{center}
\[
\begin{array}{|c| c |c |c|c|}
\hline
&&&& \cr
\phi & \sin^2 2 \theta_{solar} & \sin^2 2 \theta_{atm}
& \sin^2 2 \theta_{e3} & J \cr
\hline
&&&& \cr
0.50 & 0.6717 & 1.0 & 0 & 0\cr
&&&& \cr
0.52 & 0.7044 & 1.0 & 0 & 0\cr
&&&& \cr
0.54 & 0.7358 & 1.0 & 0 & 0\cr
&&&& \cr
0.56 & 0.7658 & 1.0 & 0 & 0\cr
&&&& \cr
0.58 & 0.7942 & 1.0 & 0 & 0\cr
&&&& \cr
0.60 & 0.8209 & 1.0 & 0 & 0\cr
&&&& \cr
\hline
\end{array}
\]
\end{center}
\caption{Results for the hyperbolic ansatz 1}
\label{table2}
\end{table}

The results for the hyperbolic ansatz 1 are given in Table \ref{table2}.
We see that depending on angle $\phi$, the quantity $\sin^2 2
\theta_{solar}$ changes, whereas $\sin^2 2 \theta_{atm}$ remain
strictly maximal at $1.0$ for all $\phi$. $U_{e3}$ is also strictly
zero for all $\phi$. Therefore this ansatz is in very good agreement
with currently favored experimental values. Note that $\phi$ is
not necessarily real. When $\phi$ takes real values so, $J=0$, the
results are given in Table \ref{table2}.

One may wonder whether $U^0$ is unitary? To answer that question
let us calculate,
\begin{equation}
{U^0}^\dagger~U^0=\pmatrix{A & B &0 \cr B & A & 0 \cr 0 & 0 & A}
\end{equation}
Where $A={\rm sech}\ ^2 \phi + \tanh^2 \phi$, $B={\rm sech}\ \phi
(\tanh \phi)^* - \tanh \phi ({\rm sech}\ \phi)^*$. One can check
analytically or using MATHEMATICA \cite{newref1-math}
that $B=0$ and $A=1$ condition is satisfied for real $\phi$.
For general complex $\phi$ we get that $A \ne 1$ and $B \ne
0$. Therefore we confine ourselves only to real $\phi$.

\noindent \underbar{HYPERBOLIC ANSATZ 2}

\begin{equation}
U=\pmatrix{
    -{\rm sech}\ \phi ~\cos \delta& \tanh \phi & {\rm sech}\ \phi~\sin
    \delta \cr
   { \sin \delta +\cos \delta~\tanh \phi \over \sqrt{2}} &
   { {\rm sech}\ \phi \over \sqrt{2}} & { \cos \delta - \sin \delta
~\tanh \phi \over \sqrt{2}} \cr
    { \sin \delta + \cos \delta ~\tanh \phi \over \sqrt{2}} &
    { {\rm sech}\ \phi \over \sqrt{2}} &
    -{\cos \delta + \sin \delta ~\tanh \phi \over \sqrt{2}}} ~~~~
\lim_{\delta \rightarrow 0} U = U^0.
\end{equation}

\begin{table}[ht]
\begin{center}
\[
\begin{array}{|c|c|c|c|c|c|}
\hline
&&&&& \cr
\phi & \delta & \sin^2 2 \theta_{solar} & \sin^2 2 \theta_{atm}
& \sin^2 2 \theta_{e3} & J \cr
\hline
&&&&& \cr
0.50 & 0.1& 0.6756 & 0.9914 & 0.0311 & 0\cr
&&&&& \cr
0.55 & 0.1& 0.7547 & 0.9899 & 0.0296 & 0\cr
&&&&& \cr
0.60 & 0.1& 0.8244 & 0.9884 & 0.0281 & 0\cr
&&&&& \cr
0.50 & 0.01& 0.6718 & 0.9999 & 0.00031 & 0\cr
&&&&& \cr
0.55 & 0.01& 0.7510 & 0.9999 & 0.00029 & 0\cr
&&&&& \cr
0.60 & 0.01& 0.8209 & 0.9998 & 0.00028 & 0\cr
&&&&& \cr
\hline
\end{array}
\]
\end{center}
\caption{Results for the hyperbolic ansatz 2.}.
\label{table3}
\end{table}

The results for the hyperbolic ansatz 2 are given in Table
\ref{table3}. We see that $U_{e3}$ is small for small $\delta$.
Due to unitarity relation of the e'th row $U_{e1}^2 + U_{e2}^2 +
U_{e3}^2=1$, this accentuation will alter the solar and atmospheric
mixing angles slightly. Therefore the atmospheric
mixing is not strictly maximal any more in this case, however
$\sin^2 2 \theta_{atm}$ remains very close to 1,
$\sin^2 2 \theta_{solar}$ remains well within the Large Mixing Angle
(LMA) solution region.

One can again question whether the matrix $U$ is unitary. The answer
is that $U$ can be factorised as,
\begin{equation}
U=U^0 \times O_{13},
\end{equation}
where,
\begin{equation}
O_{13}=\pmatrix{\cos \delta & 0 & \sin \delta \cr
0 & 1 & 0 \cr
-\sin \delta & 0 & \cos \delta}.
\end{equation}
Because $U^0$ and $O_{13}$ are unitary matrices, we find that $U$ is unitary.

Now let us discuss whether these ansatze will lead to CP violation.
For ansatz 1 and 2, let the Jarlskogian
invariant\cite{ref5-j1,ref5-j2,ref5-j3,ref5-j4,ref5-j5} given by
${\rm Im}\ [U_{e1} U_{\mu 2} U^*_{e2} U^*_{\mu_1}]$, which vanishes
if and only if CP is conserved, be denoted by $J_1$ and $J_2$. Then
we get,
\begin{eqnarray}
J_1 &=& -{1 \over 2 }{\rm Im}\ [({\rm sech}\ \phi)^2 ~((\tanh
\phi)^*)^2] \\
J_2 & = & -{1 \over 2}
{\rm Im}[
({\rm sech}\ \phi)^2~
\cos \delta~(\sin \delta + \cos \delta \tanh
\phi)^*~(\tanh \phi)^*].
\end{eqnarray}
So we see that ansatz 2 may violate CP in principle and ansatz
1 may also violate it for complex $\phi$. Because we are considering
only real $\phi$ ansatz 1 will give $J_1=0$ conserving CP. Let us take
a sample complex value
of $\delta=0.01 + 0.01~i$ along with a real $\phi=0.55$, then
we get four predictions, $\sin^2 2 \theta_{solar}=0.75$, $\sin^2
2 \theta_{atm}=0.99$, $\sin^2 2 \theta_{e3}=0.00059$ and
$J= 0.0000132$. Therefore a small but non-zero values
of $\sin^2 2 \theta_{e3}$ as well as $J$ are obtained in ansatz 2,
allowing it to violate CP. Predictions of Antatz 2 will smoothly
carry over to those of Ansatz 1 in the limit
$\delta \rightarrow 0$ where there is necessarily no CP violation.
Furthermore small values
of $\delta$ are needed to get correct solar and atmospheric mixing
in ansatz 2.

Following relationships of hyperbolic functions with circular
functions may be useful in linking our ansatze with more conventional
ansanze using circular functions.
\begin{eqnarray}
{\rm sech}\ \phi &=& \sec (i \phi) \\
\tanh \phi &=& -i \tan (i \phi)
\end{eqnarray}
For purely imaginary $\phi$ one can immediately make a correspondence
with more conventional pictures using circular functions. Because
$\phi$ need not be purely imaginary it is not immediately apparent how
to interpret this insatze in terms of simple rotation angles and
corresponding circular functions such as Sine and Cosine.

We must clarify what have we meant by the term ``Prediction''.
In any model of physical systems, if there are N measurable outputs
for (N-n) input parameters, the model is said to have `n'
``predictions''. Predictions are important from the point of view of
falsifiability. Suppose in experimental measurements we find $U_{e3}
\ne 0$, then our HYPERBOLIC ANSATZ 1 will be proven to be false. If we have N
inputs for N outputs then one can always adjust N inputs to correctly
reproduce N experimental outputs. In this case the physical model can
never be proven wrong. Therefore such physical models will not be
interesting.

To conclude, in this article we have put forward two ansatze for
lepton mixing. These ansatze use hyperbolic functions for which
arguements may either be real or complex. To preserve unitarity of
mixing matrix we should keep $\phi$ real. To the best of our
knowledge parametrizations of lepton mixing using hyperbolic
functions do not exist in literature. Present parametrization is
capable of predicting correct values for solar, atmospheric and
CHOOZ\cite{ref6-chooz1,ref6-chooz2,ref6-chooz3,ref6-chooz4,ref6-chooz5}
angles for natural values of parameters. Such parametrizations for
the lepton mixing matrix are important in the sense that they
constitute a first step toward building more complete theories of
flavor physics. From experimental point of view a good number of new
experiments such as DOUBLE-CHOOZ\cite{newref2-doublechooz}, new
reactor experiments\cite{newref3-reactor1,newref3-reactor2}, as well
as long-baseline
experiments\cite{ref7-lble1,ref7-lble2,ref7-lble3,ref7-lble4} are
either coming up or being proposed, those will be used to measure
angles including $\theta_{e3}$, where our ansatze can undergo
rigorous testing and verification.


\begin{thebibliography}{99}

\bibitem{ref1-exp1}
Super-Kamiokande Collaboration (Y. Fukuda et al.), Phys.Rev.Lett.
{\bf 81} 1562 (1998), Phys. Lett. {\bf B539} 179 (2002)

\bibitem{ref1-exp2}B.T. Cleveland et al; Astrophys.J. {\bf 496} 505 (1998)

\bibitem{ref1-exp3} SAGE Collaboration (J.N. Abdurashitov et al.), J. Exp. Theor.
Phys. {\bf 95} 181 (2002), e-Print Archive: astro-ph/0204245

\bibitem{ref1-exp4} GALLEX Collaboration (W. Hampel et al.), Phys. Lett. {\bf B447}
127 (1999)

\bibitem{ref1-exp5} SNO Collaboration (Q.R. Ahmad et al.). Phys. Rev. Lett. {\bf 89}
011301 (2002); Phys. Rev. Lett. {\bf 89} 011302 (2002)

\bibitem{ref1-exp6} KamLAND Collaboration, Phys. Rev. Lett. {\bf 90} 021802 (2003)

\bibitem{ref1-exp7} LSND Collaboration (C. Athanassopoulos et al.); Phys. Rev. Lett.
{\bf 77} 3082 (1996); Phys. Rev. Lett. {\bf 81} 1774 (1998)

\bibitem{ref2-data1}

S. Goswami, A. Bandyopadhyay, S. Choubey, Talk given at 21st
International Conference on Neutrino Physics and Astrophysics
(Neutrino 2004), Paris, France, 14-19 Jun 2004. e-Print Archive:
hep-ph/0409224

\bibitem{ref2-data2} M.C. Gonzalez-Garcia Michele Maltoni, Talk given at 5th Workshop
on Neutrino Oscillations and their Origin (NOONE2004), Tokyo, Japan,
11-15 Feb 2004. e-Print Archive: hep-ph/0406056

\bibitem{ref2-data3}M. Maltoni, T. Schwetz, M.A. Tortola, J.W.F. Valle, New J. Phys.
{\bf 6}, 122 (2004)

\bibitem{ref2-data4}K2K Collaboration (M.H. Ahn et al.), Phys. Rev. Lett. {\bf 90}
041801 (2003)

\bibitem{ref2-data5}G.L. Fogli, E. Lisi, A. Marrone, D. Montanino, A. Palazzo, Phys.
Rev. {\bf D66} 053010,2002

\bibitem{ref2-data6}R. N. Mohapatra, Pramana {\bf 63}, 1295 (2004)

\bibitem{ref2-data7} U. A. Yajnik, Pramana {\bf 63}, 1617 (2004).

\bibitem{ref3-unum1}
Zhi-zhong Xing, Int. J. Mod. Phys. {\bf A19}, 1 (2004)

\bibitem{ref3-unum2}
V. Barger, D. Marfatia, K. Whisnant, Int. J. Mod. Phys. {\bf E12}
569 (2003)

\bibitem{ref3-unum3}  G. Altarelli, F. Feruglio, New J. Phys. {\bf 6} 106 (2004)

\bibitem{ref3-unum4}
R.N. Mohapatra, UMD-PP-03-022, ICTP Lectures,  e-Print Archive:
hep-ph/0211252

\bibitem{ref3-unum5} G. Altarelli, F. Feruglio, Phys. Rept. {\bf 320} 295 (1999)

\bibitem{ref4-ansatz1}
L. Wolfenstein, Phys. Rev. {\bf D18} 958 (1978)

\bibitem{ref4-ansatz2}
A. Zee, Phys. Rev. {\bf D68}, 093002 (2003)

\bibitem{ref4-ansatz3}Zhi-zhong Xing, J. Phys. {\bf G29}, 2227 (2003)

\bibitem{ref4-ansatz4}V. D. Barger, S. Pakvasa, T. J. Weiler, K. Whisnant, Phys. Lett.
{\bf B437} 107 (1998)

\bibitem{ref4-ansatz5}R. N. Mohapatra, S. Nussinov, Phys. Lett. {\bf B441} 299 (1998)

\bibitem{ref4-ansatz6}R.N. Mohapatra, S. Nussinov, Phys. Lett. {\bf B346}, 75 (1995)

\bibitem{ref4-ansatz7}C. Giunti, Presented at 31st International Conference on High
Energy Physics (ICHEP 2002), Amsterdam, The Netherlands, 24-31 Jul
2002. Published in *Amsterdam 2002, ICHEP* 24-28, e-Print Archive:
hep-ph/0209103

\bibitem{ref4-ansatz8}W. Rodejohann, Phys. Rev. {\bf D69}, 033005 (2004)

\bibitem{ref4-ansatz9}Nan Li, Bo-Qiang Ma, Phys. Rev. {\bf D71} 017302 (2005)

\bibitem{ref4-ansatz-9-1} Nan Li, Bo-Qiang Ma, Phys. Lett. {\bf B600} 248 (2004)

\bibitem{ref4-ansatz10}V. Gupta, Xiao-Gang, Phys. Rev. {\bf D64} 117301 (2001)

\bibitem{ref4-ansatz11}D. Cocolicchio, M. Viggiano, IFUM-FT-640-99, e-Print Archive:
hep-ph/9906228

\bibitem{ref4-ansatz12}Nan Li, Bo-Qiang Ma, e-Print Archive: hep-ph/0501226

\bibitem{ref4-ansatz13}H. Fritzsch, Zhi-zhong Xing, Phys. Lett. {\bf B598}, 237 (2004)

\bibitem{ref4-ansatz13}
Osamu Yasuda, Hisakazu Minakata, e-Print Archive: hep-ph/9602386

\bibitem{ref4-ansatz14} H. Fritzsch and Z.Z. Xing, Phys. Lett. {\bf B 372}, 265 (1996).

\bibitem{ref5-j1}
C. Jarlskog, Phys. Rev. Lett. {\bf 55} 1039 (1985)

\bibitem{ref5-j2}I. Dunietz, O.W. Greenberg, Dan-di Wu, Phys. Rev. Lett. {\bf 55}
2935 (1985)

\bibitem{ref5-j3}O. Lebedev, Phys. Rev. {\bf D67} 015013 (2003)

\bibitem{ref5-j4}S. M. Bilenky, C. Giunti, W. Grimus, Phys. Rev. {\bf D58} 033001
(1998)

\bibitem{ref5-j5}J.A. Aguilar-Saavedra, J. Phys. {\bf G24} L31-L36 (1998).


\bibitem{ref6-chooz1}
M. Apollonio et al.., Eur. Phys. J. {\bf C27} 331 (2003)

\bibitem{ref6-chooz2}S. Berridge et al, e-Print Archive: hep-ex/0410081

\bibitem{ref6-chooz3} M. Goodman, e-Print Archive: hep-ph/0501206

\bibitem{ref6-chooz4}Th. Lasserre, e-Print Archive: hep-ex/0409060

\bibitem{ref6-chooz5} P. Ramond, e-Print Archive: hep-ph/0405176.


\bibitem{newref2-doublechooz}
DOUBLE-CHOOZ Collaboration, Nucl. Phys. Proc. Suppl. {\bf 145} 182
(2005)

\bibitem{newref3-reactor1}
BRAIDWOOD Collaboration, Nucl. Phys. Proc. Suppl. {\bf 149} 166
(2005)

\bibitem{newref3-reactor2}DAYA BEY Collaboration, Proceedings of 7th International Workshop
on Neutrino Factories and Superbeams (NuFact 05), hep-ph/0509041.

\bibitem{ref7-lble1}MINOS Collaboration, AIP Conf. Proc. {\bf 721} 179
(2004)

\bibitem{ref7-lble2}OPERA Collaboration, AIP Conf. Proc. {\bf 721} 231 (2004)

\bibitem{ref7-lble3}ICARUS Collaboration, Acta Phys. Polon. {\bf B34} 5385 (2003)

\bibitem{ref7-lble4}
INO Collaboration, Pramana {\bf 63} 1283 (2004).


\bibitem{newref1-math}
Mathematica, Wolfram Research, Inc, 100 Trade Center Drive
Champaign, IL 61820-7237  USA.


\end{thebibliography}
\end{document}